\documentclass[12pt,a4paper]{article}
\usepackage{epsfig}
\begin{document}
\title{The process $e^{+}e^{-}\rightarrow\nu\bar{\nu}\gamma$ in
topcolor-assisted technicolor models }

\author{Chongxing Yue$^{a}$,  Hongjie Zong$^{b}$, Wei Wang$^{a}$\\
{\small $^{a}$ Department of Physics, Liaoning Normal University,
Dalian 116029, China}\thanks{E-mail:cxyue@lnnu.edu.cn}\\
{\small $^{b}$ College of Physics and Information Engineering,}\\
\small{ Henan Normal University, Henan 453002, China } \\}
\date{\today}

\maketitle
\begin{abstract}
 In the context of topcolor-assisted technicolor (TC2) models, we
 consider the process $e^{+}e^{-}\rightarrow \nu\bar{\nu}\gamma$
 and calculate the cross section of this process at leading order.
 It is shown that the extra $U(1)$ gauge boson $Z^{\prime}$
 predicted by TC2 models can give significant contributions to the
 process $e^{+}e^{-}\rightarrow \nu_{\tau}\bar{\nu}_{\tau}\gamma$, which may be
 detected in the future high energy linear $e^{+}e^{-}$ collider
 (LC) experiments.
 \end{abstract}

 \newpage
  The cause of electroweak symmetry breaking (EWSB) and the origin
  of fermion masses are important problems of current particle
  physics. The present and next generation of colliders will help
  explain the nature of EWSB and the origin of fermion masses. The
  LHC is expected to directly probe possible new physics (NP) beyond
  the standard model (SM) up to a scale of a few $TeV$, while the high
  energy linear $e^{+}e^{-}$ collider (LC) is required to
  complement the probe of the new particles with detailed
  measurements. Furthermore, some kinds of NP predict the
  existence of new particles that would be manifested as rather
  spectacular resonance in the LC experiments, if the achievable
  centre-of-mass energy is sufficient. Even their masses exceed the
  centre-of-mass energy, it also retains an indirect sensitivity
  through a precision study the virtual corrections to the electroweak
  observables. A LC represents an ideal laboratory for studying
  this kind of NP [1].

 The  production of one or more photons and missing energy in high
 energy $e^{+}e^{-}$ collisions is a process of great interest for the
 LEP and future LC experiments. The process $e^{+}e^{-}\rightarrow \nu\bar{\nu}\gamma$
 in the SM has been successfully used for giving the number
 of light neutrino species. The events with single- and multi- photon
 final states plus missing energy play an important role in the search for the
 new phenomena of NP beyond the SM [2].

 In the context of the SM, the process $e^{+}e^{-}\rightarrow
 \nu\bar{\nu}\gamma$ has been extensively studied in leading
 order, one-loop QED, and three-loop QCD corrections [2, 3, 4].
 Recently, Ref.[5] has examined the sensitivity of this process to
 extra gauge bosons $W^{\prime}$, which arise in
 the left-right symmetric model (LRM) [6], un-unified model (UUM)
 [7], and the $KK$ model [8]. They find that the process $e^{+}e^{-}
\rightarrow \nu\bar{\nu}\gamma$ can be used to detected these new particles
 up to several $TeV$, which depends on the model, the centre-of-mass energy,
 and the assumed integrated luminosity. Furthermore, if these new
 particles are discovered at LHC or other experiments, this
 process can also be used to measure the couplings of $W^{\prime}$ to neutrinos.

In this note, we consider the process $e^{+}e^{-}\rightarrow
\nu\bar{\nu} \gamma$ in the framework of topcolor-assisted
technicolor (TC2) models [9] and calculate the contributions of
the extra $U(1)$ gauge boson $Z^{\prime}$ to this process. We find
that the gauge boson $Z^{\prime}$ can only give significant
contributes to the process $e^{+}e^{-}\rightarrow
\nu_{\tau}\bar{\nu}_{\tau} \gamma$. When the centre-of-mass
$\sqrt{s}$ is slightly larger the extra gauge boson $Z^{\prime}$
mass $M_{Z^{\prime}}$, the peak of each curve emerges. With
reasonable values of the parameters in TC2 models , the cross
section $\sigma$ of the process $e^{+}e^{-}\rightarrow
\nu_{\tau}\bar{\nu}_{\tau}\gamma$ can reach 2.4pb. Thus, this
process may be used to detect the extra $U(1)$ gauge boson
$Z^{\prime}$ and further test TC2 models in the future LC
experiments.

It is believed that there may be a common origin for EWSB and top quark mass
 generation. Much theoretical work has been carried out in connection to the
 top quark and EWSB. TC2 models [9] and flavor-universal TC2 models [10] are
 two of such examples. A common feature of such type of models is that the
 existence of extra $U(1)$ gauge bosons $Z^{\prime}$ is predicted. These new
 particles treat the third generation fermions differently from those in the
 first and second generations. The couplings of the gauge boson $Z^{\prime}$
 relevant to our calculation can be written as [11, 12]:
\begin{eqnarray}
{\cal L}_{Z^{\prime}}&=&-\frac{g_{1}}{2}[\cot\theta^{\prime}(\bar{\tau}_{L}
\gamma^{\mu}\tau_{L}+2\bar{\tau}_{R}\gamma^{\mu}\tau_{R}+\bar{\nu}_{\tau L}
\gamma^{\mu}\nu_{\tau L})-\tan\theta^{\prime}(\bar{e}_{L}
\gamma^{\mu}e_{L}+2\bar{e}_{R}\gamma^{\mu}e_{R} \nonumber  \\
&&+\bar{\nu}_{eL} \gamma^{\mu}\nu_{eL}+\bar{\mu}_{L}
\gamma^{\mu}\mu_{L}+2\bar{\mu}_{R}\gamma^{\mu}\mu_{R}+\bar{\nu}_{\mu
L} \gamma^{\mu}\nu_{\mu L})]\cdot Z_{\mu},
\end{eqnarray}
where $g_{1}$ is the ordinary hypercharge gauge coupling constant,
$\theta^{\prime}$ is the mixing angle. To obtain the top quark
condensation and not form the bottom quark condensation, there
must be $\tan\theta^{\prime}\ll 1$ [9, 11].

We can see from Eq.(1) that the extra gauge boson $Z^{\prime}$ exchange can
indeed contribute to the process $e^{+}e^{-}\rightarrow \nu\bar{\nu}\gamma$.
 However, compared to the process $e^{+}e^{-}\rightarrow \nu_{\tau}
\bar{\nu}_{\tau}\gamma$, the cross sections of the processes
$e^{+}e^{-} \rightarrow \nu_{e}\bar{\nu}_{e}\gamma$ and
$e^{+}e^{-} \rightarrow \nu_{\mu}\bar{\nu}_{\mu}\gamma$ are
suppressed by the factor $\tan^{4}\theta^{\prime}$. Thus, we can
ignore the corrections of $Z^{\prime}$ to the processes
$e^{+}e^{-}\rightarrow \nu_{\mu}\bar{\nu}_{\mu} \gamma$,
$\nu_{e}\bar{\nu}_{e}\gamma$ and only consider the contributions
of $Z^{\prime}$ to the process $e^{+}e^{-}\rightarrow
\nu_{\tau}\bar{\nu}_{\tau} \gamma$.

In the SM, the process $e^{+}e^{-}\rightarrow \nu\bar{\nu}\gamma$
proceeds at tree-level through s-channel Z exchange and t-channel
W exchange with a photon being radiated from every possible
charged particle. The relevant Feynman diagrams are shown in Fig.1
of Ref.[2]. In our calculation, we must consider the interference
effects between the electroweak gauge bosons $W$, $Z$ and extra
$U(1)$ gauge boson $Z^{\prime}$ on the cross section of the
process $e^{+}e^{-}\rightarrow \nu\bar{\nu}\gamma$. However, since
the $Z^{\prime}$ has significant contributions to the process
$e^{+}e^{-} \rightarrow \nu_{\tau}\bar{\nu}_{\tau}\gamma$ only,
the interference term relative only to the couplings
$Z\nu_{\tau}\bar{\nu}_{\tau}$ and
$Z^{\prime}\nu_{\tau}\bar{\nu}_{\tau}$. The Feynman diagrams for
the process
  $e^{+}e^{-}\rightarrow \nu_{\tau}\bar{\nu}_{\tau}\gamma$ are depicted in
Fig.1 at leading order. The main aim of this paper is to calculate
the contributions of TC2 models to the process
$e^{+}e^{-}\rightarrow \nu\bar{\nu}\gamma$ and see whether the
 $Z^{\prime}$ can be detected in the future LC experiments via this process. If the extra $U(1)$ gauge
boson $Z^{\prime}$ is indeed observed, it will be important to
study high order effects.

\begin{figure}[htb]
\vspace{-3.5cm}
\begin{center}
\epsfig{file=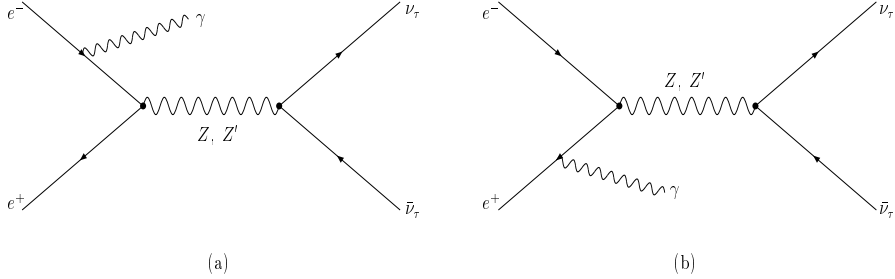,width=350pt,height=520pt} \vspace*{-9cm}
\hspace{5mm} \caption{The Feynman diagrams contribute to the
process  $e^{+}e^{-}\rightarrow \nu_{\tau}\bar{\nu}_{\tau}\gamma$
 at leading order.}
 \label{ee}
\end{center}
\end{figure}

Let M denote the sum of the amplitudes shown in Fig.1. Using Eq.(1) and
other relevant Feynman rules, squaring the helicity amplitudes and summing
over the final state helicities, the spin-averaged unpolarized $|M|^{2}$ for
the process  $e^{+}(p_{+})+e^{-}(p_{-})\rightarrow \nu(q_{-})+\bar{\nu}
(q_{+})+\gamma(k)$ can be written as :
 \begin{eqnarray}
|M|^{2}&=&\frac{(4\pi)^{3}\alpha^{3}}{2S_{W}^{4}C_{W}^{4}}
\frac{s^{\prime}}{k_{+}k_{-}}\{[\frac{(S_{W}^{2}-\frac{1}{2})^{2}}{Z^{2}}+
\frac{S_{W}^{4}}{4Z^{\prime 2}}+\frac{(S_{W}^{2}-\frac{1}{2})S_{W}^{2}}
{Z\cdot Z^{\prime}}](u^{2}+u^{\prime 2}) \nonumber  \\
&&+[\frac{1}{Z^{2}}+\frac{1}{Z^{\prime 2}}+\frac{2}{Z\cdot Z^{\prime}}]
S_{W}^{4}(t^{2}+t^{\prime 2})\},
\end{eqnarray}
with
\begin{equation}
s^{\prime}=(q_{+}+q_{-})^{2},\hspace{4mm} u=(p_{+}-q_{-})^{2},\hspace{4mm} u^{\prime}=(p_{-}-q_{+})^{2},
\end{equation}
\begin{equation}
t=(p_{+}-q_{+})^{2},\hspace{4mm} t^{\prime}=(p_{-}-q_{-})^{2},\hspace{4mm} k_{\pm}=2p_{\pm}k,\hspace{6mm}
\end{equation}
 \begin{equation}
Z=s^{\prime}-M_{Z}^{2}+iM_{Z}\Gamma_{Z},\hspace{8mm}
Z^{\prime}=s^{\prime}-M_{Z^{\prime}}^{2}+iM_{Z^{\prime}}\Gamma_{Z^{\prime}}.
\end{equation}
Where $\alpha$ is the electromagnetic coupling constant,
$S_{W}=\sin\theta_{W}$ which $\theta_{W}$ is the Weinberg angle.
$\Gamma_{Z}$ and $\Gamma_{Z^{\prime}}$ are the total decay widths
of the gauge bosons $Z$ and $Z^{\prime}$, respectively. The decay
width $\Gamma_{Z^{\prime}}$ is dominated by the decay models
$t\bar{t}$, $b\bar{b}$. In the following numerical estimation, we
will take [11] :
\begin{equation}
\Gamma_{Z^{'}}\approx
\frac{g_{1}^{2}\cot^{2}\theta^{\prime}}{12\pi}M_{Z^{\prime}}=\frac{1}{3}M_{Z^{\prime}},
\end{equation}
which corresponds to $\tan^{2}\theta^{\prime}=0.01$.

The limits on the masses of the extra $U(1)$ gauge bosons $Z^{'}$
can be obtained via studying their effects on various experimental
observables [12,13]. For example, Ref.[14] has shown that
$B\overline{B}$ mixing provides stronger lower bounds on the mass
of $Z^{'}$ predicted by TC2 models, one must has $M_{Z^{'}}>
6.8\hspace{2mm} TeV(9.6 \hspace{2mm} TeV)$ if ETC does(does  not)
contribute to the $cp-$violation parameter $\varepsilon$.
Recently, Ref.[15] has restudied the bound placed by the
eletroweak measurement data on the $Z^{'}$ mass. They find that
$Z^{'}$ predicted by TC2 models must be heavier than $1
\hspace{2mm} TeV$. As numerical estimation, we will take
$M_{Z^{\prime}}$ as a free parameter and assume it in the range of
$1\sim 3 \hspace{2mm} TeV$.

For the process $e^{+}e^{-}\rightarrow
\nu_{\tau}\bar{\nu}_{\tau}\gamma$, the signal is a detected,
energetic photon. The kinematic variable of interest are the
photon's energy $E_{\gamma}$ and its angle relative to the
incident electron, $\theta_{\gamma}$, which are both defined in
the $e^{+}e^{-}$ centre-of-mass frame [5]. The doubly differential
cross section of this process can be written as :
\begin{equation}
\frac{d\sigma}{dE_{\gamma}d\cos\theta_{\gamma}}=\frac{E_{\gamma}}{2s}
\frac{1}{(4\pi)^{4}}\int_{0}^{\pi}d\theta\sin\theta\int_{0}^{2\pi}d\varphi
 |M|^{2},
\end{equation}
where $\sqrt{s}$ is the centre-of-mass energy of LC experiments,
$\theta$ and $\varphi$ are the polar and azimuthal angles of
$q_{+}$, in a frame where $q_{+}$ and $q_{-}$ are back-to-back.
\begin{figure}[htb]
\vspace{-0.9cm}
\begin{center}
\epsfig{file=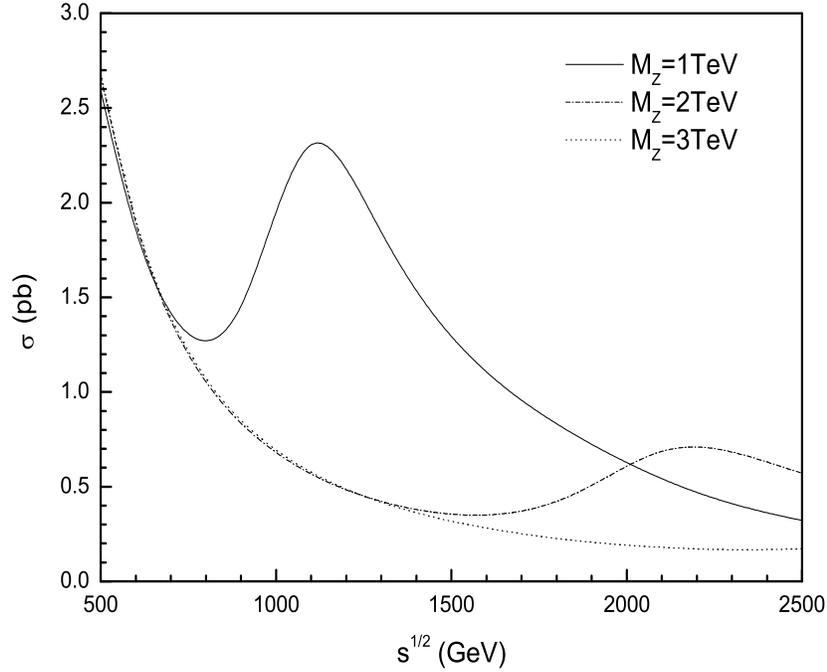,width=350pt,height=300pt} \vspace*{-1.32cm}
 \caption{The cross section $\sigma$ of the process
         $e^{+}e^{-}\rightarrow \nu_{\tau}\bar{\nu}_{\tau}\gamma$
         versus $\sqrt{s}$ for $M_{Z^{\prime}}=1 \hspace{2mm} TeV$, $2 \hspace{2mm} TeV$, and $3 \hspace{2mm} TeV$.}
 \label{fig.2}
\end{center}
\end{figure}
\vspace*{0cm}

To obtain numerical results, we take the SM parameters as
$S_{W}^{2}=0.2315$, $\alpha=\frac{1}{128.8}$, $M_{Z}=91.2
\hspace{2mm} GeV$, and $\Gamma_{Z}=2.495 \hspace{2mm} GeV$[16].
Kinematically, the maximum allowed value for $E_{\gamma}$ is
$\frac{\sqrt{s}}{2}$. In our numerical integral, we take the
kinematic variables in the ranges : $10 \hspace{2mm} GeV\leq
E_{\gamma}\leq \frac{\sqrt{s}}{2}$ and $10^{o}\leq
\theta_{\gamma}\leq 170^{o}$ [5].

The cross section $\sigma$ of the process $e^{+}e^{-}\rightarrow
\nu_{\tau} \bar{\nu}_{\tau}\gamma$ is plotted in Fig.2 as a
function of $\sqrt{s}$ for three values of $M_{Z^{\prime}}$:
$M_{Z^{\prime}}=1 \hspace{2mm} TeV$, $2 \hspace{2mm} TeV$ and $3
\hspace{2mm} TeV$. From Fig.2 we can see that the cross section
$\sigma$ increases as $M_{Z^{\prime}}$ and $\sqrt{s}$ decreasing
in most of the parameter space. However, the resonance peak
emerges for $\sqrt{s}$ slightly above the $Z^{\prime}$ mass
$M_{Z^{\prime}}$. For $\sqrt{s}=1100 \hspace{2mm} GeV$ and
$M_{Z^{\prime}}=1000 \hspace{2mm} GeV$, the value of the cross
section $\sigma$ is 2.41 pb. As long as $\sqrt{s}\leq 2
\hspace{2mm} TeV$ and $M_{Z^{\prime}}\leq 3 \hspace{2mm} TeV$,
$\sigma$ is larger than 0.19 pb.

\begin{figure}[htb]
\vspace*{-0.5cm}

\begin{center}
\vspace*{0cm}
 \epsfig{file=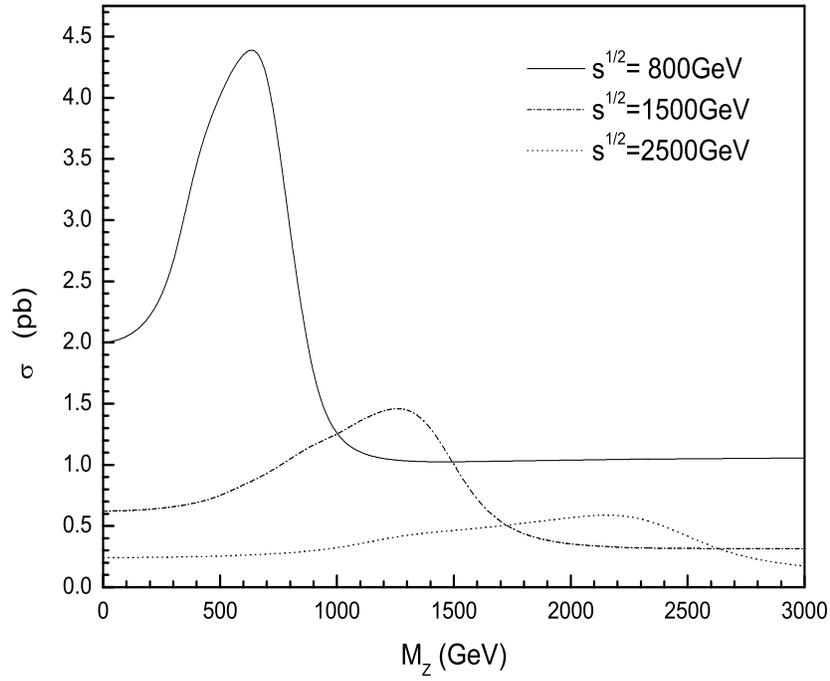,width=350pt,height=300pt}
\vspace*{-1cm} \caption{The cross section $\sigma$ as a function
of $M_{Z^{\prime}}$ for
         $\sqrt{s}=800 \hspace{2mm} GeV$, $1500 \hspace{2mm} GeV$ and $2500 \hspace{2mm} GeV$.}
\label{fig.3}
\end{center}
\end{figure}

\vspace*{-0.5cm}

To see the effect of the extra $U(1)$ gauge boson $Z^{\prime}$
mass $M_{Z^{\prime}}$ on the cross section $\sigma$, we plot
$\sigma$ versus $M_{Z^{\prime}}$ for $\sqrt{s}=800 \hspace{2mm}
GeV$, $1500 \hspace{2mm} GeV$ and $2500 \hspace{2mm} GeV$ in
Fig.3. From Fig.3 we can see that the cross section $\sigma$ is
suppressed by large $Z^{\prime}$ mass $M_{Z^{\prime}}$. For
$\sqrt{s}=1500 \hspace{2mm} GeV$, $M_{Z^{\prime}}=1300
\hspace{2mm} GeV$, the value of $\sigma$ can reach 1.48 pb. If we
assume $M_{Z^{\prime}}=700 \hspace{2mm} GeV$, then we have
$\sigma=4.43 pb$ for $\sqrt{s}=800 \hspace{2mm} GeV$.

The process $e^{+}e^{-}\rightarrow \nu\bar{\nu}\gamma$, which has
a detected, energetic photon final state and missing energy plays
an important role in the search for NP. It has been extensively
studied in the SM and some specific models, such as LRM, UUM and
$KK$ models. However, it is very little to study this process in
the context of dynamical models of EWSB. In this note, we discuss
this process in the framework of TC2 models. We find that, since
the extra gauge boson $Z^{\prime}$ treats the third generation
fermions differently to those of the first and second generation
fermions and couples preferentially to the third generation
fermions,  the cross section $\sigma$ of the process
$e^{+}e^{-}\rightarrow \nu_{\tau}\bar{\nu}_{\tau}\gamma$ can be
significantly enhanced.  With reasonable values of the parameters
in TC2 models , $\sigma$ can reach 2.4 pb, which may be detected
in the future LC experiments. Thus, the process
$e^{+}e^{-}\rightarrow \nu_{\tau}\bar{\nu}_{\tau}\gamma$ can be
used to probe the possible signals of $Z^{\prime}$ and further
unravel the TC2 models.

\vspace{.5cm}
\noindent{\bf Acknowledgments}

Chongxing Yue thanks the Abdus Salam International Centre for
Theoretical Physics (ICTP) for partial support. This work was
supported in part by the National Natural Science Foundation of
China (90203005).


\null

\end{document}